%% file: main.tex
\documentclass[sigconf]{acmart}
\AtBeginDocument{%
  \providecommand\BibTeX{{%
    \normalfont B\kern-0.5em{\scshape i\kern-0.25em b}\kern-0.8em\TeX}}}

\copyrightyear{2024}
\acmYear{2024}
\setcopyright{acmlicensed}\acmConference[RecSys '24]{18th ACM Conference
on Recommender Systems}{October 14--18, 2024}{Bari, Italy}
\acmBooktitle{18th ACM Conference on Recommender Systems (RecSys '24),
October 14--18, 2024, Bari, Italy}
\acmDOI{10.1145/3640457.3688170}
\acmISBN{979-8-4007-0505-2/24/10}

\usepackage{subcaption}
\usepackage{enumitem}

\begin{document}

\title{CAPRI-FAIR: Integration of Multi-sided Fairness in Contextual POI Recommendation Framework}


\author{Francis Zac dela Cruz}
\orcid{0009-0005-5790-3379}
\email{f.dela_cruz@student.unsw.edu.au}
\author{Flora D. Salim}\authornote{Corresponding author}
\orcid{0000-0002-1237-1664}
\email{flora.salim@unsw.edu.au}
\affiliation{%
  \institution{School of Computer Science and Engineering,\\ University of New South Wales }
    \city{Sydney}
  \state{NSW}
  \country{Australia}
}

\author{Yonchanok Khaokaew}
\orcid{0000-0003-4297-6274}
\email{y.khaokaew@unsw.edu.au}
\affiliation{%
  \institution{School of Computer Science and Engineering,\\ University of New South Wales }
    \city{Sydney}
  \state{NSW}
  \country{Australia}
}


\author{Jeffrey Chan}
\orcid{0000-0002-7865-072X}
\email{jeffrey.chan@rmit.edu.au}
\affiliation{
    \institution{School of Computing Technologies\\
    RMIT University}
    \city{Melbourne}
    \country{Australia}}


\begin{abstract}
Point-of-interest (POI) recommendation considers spatio-temporal factors like distance, peak hours, and user check-ins. Given their influence on both consumer experience and POI business, it's crucial to consider fairness from multiple perspectives. Unfortunately, these systems often provide less accurate recommendations to inactive users and less exposure to unpopular POIs. This paper develops a post-filter method that includes provider and consumer fairness in existing models, aiming to balance fairness metrics like item exposure with performance metrics such as precision and distance. Experiments show that a linear scoring model for provider fairness in re-scoring items offers the best balance between performance and long-tail exposure, sometimes without much precision loss. Addressing consumer fairness by recommending more popular POIs to inactive users increased precision in some models and datasets. However, combinations that reached the Pareto front of consumer and provider fairness resulted in the lowest precision values, highlighting that tradeoffs depend greatly on the model and dataset.
\end{abstract}





\maketitle

\section{Introduction}
\label{sec:intro}

\input{sections/introduction}




\section{Modelling the Task and the Fairness Factors}
\label{sec:modelling}

\input{sections/problem_definition}

\vspace{-5px}
\section{Experiment Setup}
\label{sec:experiment}

\input{sections/methodology}

\section{Results and Discussion}
\label{sec:results}

\input{sections/results}

\vspace{-3px}
\section{Conclusion}
\label{sec:conclusion}

\input{sections/conclusion}

\section*{ACKNOWLEDGMENTS}
This research is partly funded by the Australian Research Council Centre of Excellence for Automated Decision-Making and Society (ADM+S) (CE200100005).
\clearpage
\balance
\bibliographystyle{ACM-Reference-Format}
\bibliography{main}

\clearpage

\newpage
\appendix\label{sec:aux}
\section{Auxiliary}
\input{sections/aux}

\end{document}

%% file: sections/introduction.tex
Recommender systems are pivotal in digital environments, tailoring content across platforms like Spotify, YouTube, TikTok, and Facebook. Specifically, point-of-interest (POI) recommender systems suggest locations such as restaurants and stores, emphasizing geographical proximity as a crucial factor \cite{p10-ye2011exploiting}. While these systems significantly enhance user experience, the critical issue of fairness within these algorithms has gained substantial attention. Research in machine learning fairness has evolved, addressing various definitions and the multifaceted nature of biases in recommender systems \cite{p5-ekstrand2022fairness, p6-mehrabi2021survey}.
Existing work \cite{p1-burke2018balanced, p7-abdollahpouri2020multistakeholder} has begun to address multi-stakeholder recommendation frameworks that consider the needs of both consumers and providers. Consumers benefit most from recommendations that accurately reflect their interests and intentions, while providers gain from equitable exposure of their items. Despite these advancements, current models often struggle to balance these interests effectively without sacrificing the quality of recommendations or exacerbating existing biases.

Our research targets this gap by integrating a `fairness factor' into POI recommender systems through post-processing techniques adaptable to various underlying models. This approach aims to refine the balance between provider and consumer interests, often overlooked in conventional systems. Additionally, we explore multi-objective fairness in POI recommender systems. This area has seen limited investigation, particularly in how different fairness objectives can be simultaneously optimized without detriment to the overall system performance \cite{p47-patro2020fairrec, p15-naghiaei2022cpfair}.

This paper contributes to the field by proposing a methodology that reconciles the often conflicting objectives of different stakeholders within the POI recommendation context. By doing so, we aim to push the boundaries of what is currently possible in fairness-oriented recommender systems, providing a template for future research to build upon \cite{p2-patro2020towards, p42-merinov2023sustainability}. We pose the following research questions to guide our investigation:
\begin{itemize}[nosep, topsep=0pt, itemsep=0pt, leftmargin=15pt]
    \item[] \textbf{RQ1}: How does the scoring model for the provider fairness factor affect the recommender system's performance?
    \item[] \textbf{RQ2}: What are the impacts of the provider fairness factor on performance, fairness, and distance metrics compared to the consumer fairness factor?
    \item[] \textbf{RQ3}: How does the recommender system perform when integrating provider and consumer fairness factors?
\end{itemize}

Aligned with these questions, our investigation will test several hypotheses:
\begin{itemize}[nosep, topsep=0pt, itemsep=0pt, leftmargin=15pt]
    \item[] \textbf{H1}: The exposure model has an impact on the system's performance.
    \item[] \textbf{H2}: The provider fairness factor improves long-tail item exposure and the consumer fairness factor enhances precision for inactive users, both without compromising accuracy.
    \item[] \textbf{H3}: Integrating both provider and consumer fairness factors potentially improves GCE fairness metrics.
\end{itemize}

To address these questions, we explore the complexities of designing fairness-oriented systems. The literature supports various fairness definitions, from Aristotle's principle of treating equals equally, to more nuanced considerations of group and individual fairness in machine learning contexts \cite{p46-dwork2012awareness, p5-ekstrand2022fairness}. The spatial dimensions of POI recommendations, which consider the proximity of recommended sites to the user, add another layer of complexity, especially given the often sparse data on user interactions with potential POIs \cite{p48-levandoski2012lars, p10-ye2011exploiting}.
This study does not only address theoretical aspects of fairness and performance but also practical implementations, examining existing methodologies and proposing new ways to balance consumer and provider interests within the framework of POI recommender systems. By integrating considerations of geographical closeness, social connections, and categorical preferences, we aim to formulate a comprehensive system that aligns with contemporary needs for fairness and utility in digital recommendations.

%% file: sections/problem_definition.tex
\textbf{Problem Statement}\\
This project aims to develop a recommender system model that incorporates a user $u \in U$, a point-of-interest (POI) $p \in P$, and contextual factors $C(u, p)$ (e.g., distance, time, social connections), all of which are normalized to ensure values range between 0 and 1. Additionally, it integrates consumer and provider fairness through abstract functions $F_c(u)$ and $F_p(u)$, which also range from 0 to 1 for consumer and provider fairness, respectively. The model, denoted by $\hat{M}$, combines these contexts to produce personalized POI recommendations. Fairness is incorporated into this framework by adjusting the ranking score as follows:
\begin{equation}
\mathbf{M}(u, p) = \hat{M}(u, p, C(u, p)) + \alpha \cdot F_p(u) + \beta \cdot F_c(u)
\end{equation}

where the tunable parameters \(\alpha\) and \(\beta\) modulate the impact of fairness considerations and ensure the score \(\mathbf{M}(u, p)\) remains within the range of 0 to 1. Different configurations will be evaluated to illustrate their effects on system performance and fairness, aiming to identify optimal balances in a Pareto front.

In this front, we propose a framework CAPRI-FAIR\footnote{https://github.com/cruiseresearchgroup/CAPRI-FAIR}, where we extend a pre-existing framework for context-aware POI recommendation and incorporate fairness factors. We also add evaluations that consider user and item fairness.
\vspace{1.5mm}
\\
\textbf{Consumer Fairness}\\
Consumer fairness addresses the disparity between active and inactive users on the platform, distinguished by their frequency of check-ins. Active users benefit from more personalized recommendations due to their extensive interaction data, enhancing profile accuracy. Conversely, for inactive users, who may need a basic understanding of popular options to start exploring, a strategy of recommending widely popular POIs proves beneficial. This approach aligns with findings from Rahmani \cite{p31-rahmani2022unfairness}, indicating that less active users frequently attend popular POIs.

We propose recommending nearby, popular POIs to inactive users, defining "nearby" as POIs within the closest 20\% of distances to any previously visited POI \cite{p50-werneck2021points}. This method focuses on enhancing recommendation precision for inactive users without altering the accuracy of active ones.
\vspace{1.5mm}
\\
\textbf{Provider Fairness}\\
Each POI has an associated popularity, indicated by user check-ins, which follows a power law distribution. Most check-ins are concentrated among a few POIs, while a large majority fall into the `long tail' with minimal exposure. This distribution characteristic underscores the challenge of achieving fairness among providers, as most POIs receive scant attention. To address this, we model popularity with a power law, where the score of a POI is inversely proportional to its check-in count, helping to elevate less popular POIs. For a given number of $x$ check-ins, the estimated number of POIs $y$ with that popularity is expressed inline as $ y = w_0 \cdot x^{w_1}$. This relationship undergoes a log-log transformation to fit a linear regression, where parameters are determined using L2 regularization to manage data skew, with a regularization factor of 10.0 ensuring stable predictions. In addition to the power law model, we explore linear and logistic regression models for comparison. The linear model does not scale down exposure as sharply with increasing popularity, offering a more gradual adjustment for moderately popular POIs. In contrast, logistic regression treats slight increases in popularity more harshly, impacting even moderately popular POIs significantly. To assess provider fairness, we measure each item's exposure across recommendations, where each POI is assumed to represent a unique provider. Exposure ($E$) is quantified using an attention function \( a(p, P_R) \), which, for simplicity, is binary in our experiments, indicating whether an item \( p \) is present or absent in the recommendation list. The exposure of an item \( p \) is defined as $ (E_p = \sum_{P_R \subset P} a(p, P_R))$, aligning with methods in related work that scale exposure based on item visibility or rank within the list, facilitating comparison of different ranking algorithms' impact on provider fairness.

%% file: sections/methodology.tex
\textbf{Datasets}\\
We employ the Gowalla\cite{cho2011friendship} and Yelp\cite{liu2017experimental} datasets, recognized for their comprehensive feature sets which facilitate model comparisons \cite{p50-werneck2021points, p66-rahmani2022role}. Following standard preprocessing, we remove POIs visited by fewer than 10 users and users who have visited fewer than 10 POIs. After this filtering process, the dataset includes 7135 users for Yelp and 5628 users for Gowall. We divide each user's sequence of POI visits into 70\% training, 20\% validation, and 10\% testing, maintaining the chronological order to simulate real-world scenarios and ensure robust evaluations of our recommendation system. Users are categorized as `active' or ``inactive' based on their activity, with the top 20\% being active. This is based on the defaults given by the original CAPRI framework \cite{p70-tourani2023capri} and the Pareto principle (i.e., where 20\% of users account for 80\% of check-ins). Items are similarly divided into `short-head' and `long-tail' groups based on the top 20\% of popularity metrics.
\begin{table*}[ht!]
\small
\centering
\caption{Precision and Long-tail Exposure by Provider Fairness Weights and Exposure Models in CAPRI-FAIR framework.
}
\vspace{-7.5pt}

\resizebox{0.87\textwidth}{!}{
\begin{tabular}{|c c|c c c c |c c c c |c c c c |c c c c |}
    \hline
    \multicolumn{2}{|c|}{\textbf{Dataset}}&\multicolumn{8}{c|}{\textbf{Yelp}}&\multicolumn{8}{c|}{\textbf{Gowalla}} \\ \hline \hline
     \textbf{Model} & \textbf{Exposure Model} & \multicolumn{4}{c|}{Precision} & \multicolumn{4}{c|}{Long-tail Exposure}  & \multicolumn{4}{c|}{Precision} & \multicolumn{4}{c|}{Long-tail Exposure} \\
    &  & 0.25 & 0.50 & 0.75 & 1  & 0.25 & 0.50 & 0.75 & 1& 0.25 & 0.50 & 0.75 & 1  & 0.25 & 0.50 & 0.75 & 1  \\
    \hline\hline
     GeoSoCa & Linear  & 0.0134 & \textbf{0.0117} & \textbf{0.0111} & \textbf{0.0107}  & \textbf{3.4467} & 4.1525 & 4.5026 & 4.6888 & 0.0300 & \textbf{0.0287} & \textbf{0.0274} & \textbf{0.0266}  & 0.9796 & 1.0734 & 1.1626 & 1.2465 \\
     & Logistic  & \textbf{0.0173} & 0.0089 & 0.0073 & 0.0073  & 3.0696 & \textbf{4.5693} & \textbf{4.6969} & \textbf{4.6970} & \textbf{0.0316} & 0.0182 & 0.0140 & 0.0132 &   1.2142 & \textbf{1.9263} & \textbf{2.0336} & 2.0459 \\
     & PowerLaw  & 0.0137 & 0.0109 & 0.0096 & 0.0081  & 3.1636 & 3.5278 & 2.5025 & 1.8429 & 0.0277 & 0.0242 & 0.0222 & 0.0203  & \textbf{1.3231} & 1.8073 & 2.0207 & \textbf{2.0636}\\
    \hline
     LORE & Linear  & 0.0179 & \textbf{0.0147} & \textbf{0.0132} & \textbf{0.0124}  & 2.7975 & 3.6141 & 4.1361 & 4.4426 & \textbf{0.0427} & \textbf{0.0405} & \textbf{0.0389} & \textbf{0.0378}  & 0.9356 & 1.0757 & 1.1958 & 1.2961\\
    & Logistic  & 0.0183 & 0.0096 & 0.0074 & 0.0072  & \textbf{3.4217} & \textbf{4.8180} & \textbf{4.9503} & \textbf{4.9565} & 0.0283 & 0.0149 & 0.0127 & 0.0125  & \textbf{1.6326} & \textbf{2.0425} & \textbf{2.0815} & \textbf{2.0842}\\
    & PowerLaw  & \textbf{0.0196} & 0.0136 & 0.0106 & 0.0082  & 2.2284 & 2.1647 & 1.8887 & 1.5768& 0.0392 & 0.0260 & 0.0191 & 0.0156 & 1.2983 & 1.6527 & 1.7724 & 1.7771  \\
    \hline
    USG & Linear  & \textbf{0.0314} & 0.0278 & \textbf{0.0228} & \textbf{0.0186}  & 0.0341 & 0.5424 & 1.8214 & 3.0765 & 0.0504 & \textbf{0.0477} & \textbf{0.0446} & \textbf{0.0424}  & 0.1056 & 0.1567 & 0.2523 & 0.3689 \\
    & Logistic & 0.0296 & 0.0257 & 0.0123 & 0.0042 & \textbf{0.0461} & \textbf{1.1532} & \textbf{4.5227} & \textbf{5.6810} & 0.0500 & 0.0334 & 0.0195 & 0.0103 & \textbf{0.2319} & \textbf{1.4315} & \textbf{2.0190} & \textbf{2.2654} \\
    & PowerLaw  & 0.0300 & \textbf{0.0291} & 0.0152 & 0.0036  & 0.0209 & 0.2791 & 2.9418 & 3.9288 & \textbf{0.0525} & 0.0468 & 0.0293 & 0.0158  & 0.1513 & 0.7876 & 1.6531 & 2.0789 \\
    \hline\hline

\end{tabular}

}
\label{table:prec-ltexp-exposuremodel-yelp}
\vspace{-8pt}
\end{table*}
\vspace{1.5mm}
\\
\textbf{Evaluation Metrics}\\
For evaluation, we primarily use Precision@k as our accuracy metric, which assesses if the target POI is among the top $k$ entries in the recommendation list—suitable for scenarios with sparse hits.

We measure both Precision and Exposure across different groups to evaluate user and item fairness. For fairness assessment, we employ the Generalized Cross-Entropy (GCE) with Pearson's $\chi^2$ measure, noted for its robustness to outliers \cite{p31-rahmani2022unfairness, p55-deldjoo2021flexible}. This measure quantifies item exposure disparities, focusing on long-tail exposure improvements. For consumer fairness, recommendation gain for each user is defined based on the presence of recommended POIs in their previous visits. Similarly, provider fairness is evaluated by whether POIs appear in recommendation lists. Each POI's latitude and longitude are used to compute distances, and the user's position is averaged from the coordinates of their previously visited POIs. Distances are then calculated using the standard geographical distance formula, ensuring accurate measurement of proximity.
\vspace{1.5mm}
\\
\textbf{Benchmarking and Experimentation}\\
Our model is benchmarked against three widely used POI recommenders \cite{p70-tourani2023capri, p50-werneck2021points, yu2020category, p29-rahmani2022exploring}: USG \cite{p10-ye2011exploiting}, GeoSoCa \cite{zhang2015geosoca}, and LORE \cite{zhang2014lore}. USG integrates user preferences, social influence, and geographic data. GeoSoCa leverages geographical, social, and categorical correlations, while LORE combines geographic, temporal, and social contexts. These methods are chosen for their robust integration of diverse contextual factors crucial for effective POI recommendations. Future evaluations may extend to include graph-based, deep learning-based, and generative AI-based methods for a more comprehensive comparison. Consistent datasets are used as per \cite{p70-tourani2023capri} to maintain comparability.

Our experiments systematically evaluate the effects of fairness factors on model performance and fairness metrics. We assess three exposure models—power law, linear, and logistic—across \(\alpha\) values from 0 to 1, analyzing their impact on precision and long-tail exposure at \(k = 10\). Both provider and consumer fairness factors are explored for their influence on item exposure and precision for inactive users. We employ the Kruskal-Wallis test to assess differences among the three exposure models starting at \(\alpha = 0.25\). Pairwise Mann-Whitney U tests further analyze specific model comparisons, enhancing our evaluation of each model's impact.

%% file: sections/results.tex
\begin{table*}[h!]
\centering

\caption{Precision, GCE, and Mean Median Distance Metrics for Different Weights of Providers and Consumers, Yelp Dataset.}
\vspace{-7.5pt}
\resizebox{0.87\textwidth}{!}{
    \begin{tabular}{|c c||c c c|c c c|c c c|c c c|}
        \hline
        Model & Fairness Weights & \multicolumn{3}{c|}{Precision} & \multicolumn{3}{c|}{Item fairness GCE} & \multicolumn{3}{c|}{User fairness GCE} & \multicolumn{3}{c|}{Mean median distance} \\
        & $(\alpha, \beta)$ & @5 & @10 & @20 & @5 & @10 & @20 & @5 & @10 & @20 & @5 & @10 & @20 \\
        \hline\hline
        GeoSoCa 
         & (0.0, 0.5) & 0.0291 & 0.02537 & 0.02159 & -5.89764 & -4.54826 & -3.34275 & -0.09071 & -0.09333 & -0.09345 & 100.33198 & 99.73246 & 99.80617 \\
         & (0.0, 1.0) & \textbf{0.02949} & \textbf{0.02562} & \textbf{0.02195} & -9.9195 & -8.07117 & -6.32284 & -0.08641 & -0.09007 & -0.08808 & 102.35795 & 102.48056 & 101.51285 \\
         & (0.25, 0.25) & 0.0169 & 0.01598 & 0.01399 & -0.6379 & -0.56611 & -0.49125 & -0.04459 & -0.03675 & -0.04917 & 98.60541 & 97.4856 & 97.20788 \\
         & (0.5, 0.0) & 0.01345 & 0.01173 & 0.0108 & \textbf{-0.00053} & \textbf{-0.00334} & \textbf{-0.00759} & -0.06675 & -0.06383 & -0.06583 & \textbf{96.86406} & \textbf{96.87758} & \textbf{97.11985} \\
         & (0.5, 0.5) & 0.01603 & 0.01453 & 0.01282 & -0.57984 & -0.52849 & -0.46714 & \textbf{-0.03134} & \textbf{-0.02418} & \textbf{-0.03132} & 99.0131 & 98.98704 & 98.47554 \\
         & (1.0, 0.0) & 0.01233 & 0.01071 & 0.00955 & -0.06374 & -0.03787 & -0.02046 & -0.03798 & -0.06055 & -0.06478 & 97.4033 & 97.37404 & 97.60246 \\
         \hline
        LORE   & (0.0, 0.5) & 0.02826 & 0.02572 & 0.02305 & -3.19637 & -3.20054 & -3.06965 & -0.08243 & -0.08841 & -0.08695 & 98.88016 & 98.2366 & 98.32551 \\
         & (0.0, 1.0) & \textbf{0.02929} & \textbf{0.02657} & \textbf{0.02341} & -6.95594 & -6.77278 & -6.08884 & -0.07205 & -0.07841 & -0.0821 & 100.45197 & 99.56773 & 99.89955 \\
         & (0.25, 0.25) & 0.02105 & 0.01964 & 0.01687 & -0.58276 & -0.57503 & -0.53311 & -0.09777 & -0.09092 & -0.0879 & 97.48547 & 96.56786 & 96.24566 \\
         & (0.5, 0.0) & 0.01606 & 0.0147 & 0.01261 & -0.07092 & -0.05139 & -0.0277 & -0.1151 & -0.10619 & -0.11235 & \textbf{96.45963} & \textbf{95.95741} & \textbf{95.6067} \\
         & (0.5, 0.5) & 0.01844 & 0.01755 & 0.01499 & -0.5067 & -0.50162 & -0.47727 & \textbf{-0.06978} & \textbf{-0.0547} & \textbf{-0.05941} & 97.19982 & 96.48961 & 95.97415 \\
         & (1.0, 0.0) & 0.01407 & 0.0124 & 0.01064 & \textbf{-0.0} & \textbf{-0.00197} & \textbf{-0.01268} & -0.0963 & -0.0968 & -0.10213 & 96.85222 & 96.23217 & 95.69327 \\
         \hline
        USG          & (0.0, 0.5) & 0.03324 & 0.02882 & 0.02424 & -595.83844 & -552.20543 & -267.68722 & -0.06434 & -0.07915 & -0.09065 & 117.46961 & 115.80121 & 116.1397 \\
         & (0.0, 1.0) & 0.03263 & 0.02804 & 0.02374 & -707.64802 & -665.99347 & -419.15168 & -0.06917 & -0.08743 & -0.09763 & 120.61967 & 119.56118 & 118.70201 \\
         & (0.25, 0.25) & \textbf{0.03518} & \textbf{0.031} & \textbf{0.02639} & -182.2632 & -124.02658 & -67.06192 & -0.05818 & -0.08549 & -0.09072 & 112.82197 & 113.71123 & 113.15953 \\
         & (0.5, 0.0) & 0.03083 & 0.02783 & 0.02346 & -3.26253 & -2.88418 & -2.31393 & -0.05556 & -0.06852 & -0.07835 & 112.54663 & 113.1397 & 113.10282 \\
         & (0.5, 0.5) & 0.0317 & 0.0284 & 0.02397 & -13.47004 & -11.9686 & -9.91389 & -0.04953 & -0.06343 & -0.07234 & \textbf{112.05213} & 113.48431 & 113.014 \\
         & (1.0, 0.0) & 0.02094 & 0.01857 & 0.01501 & \textbf{-0.16983} & \textbf{-0.13466} & \textbf{-0.09566} & \textbf{-0.02918} & \textbf{-0.04137} & \textbf{-0.05436} & 112.63371 & \textbf{112.67297} & \textbf{110.77933} \\
        \hline
    \end{tabular}

}
\label{table:metrics-with-tradeoffs-yelp}
\vspace{-5pt}
\end{table*}

The benchmark models were initially run without modifications, then progressively with increased provider fairness factor $\alpha$ coefficients to assess the impact delineated in \textbf{H1}. The first assessment focused on the effect of the provider fairness model on precision and long-tail exposure metrics, which it aims to improve. Results presented in Table \ref{table:prec-ltexp-exposuremodel-yelp} highlight the outcomes of these evaluations.

\begin{figure}[ht!]
\vspace{-5px}
 \centering
            \includegraphics[width=0.90\linewidth]{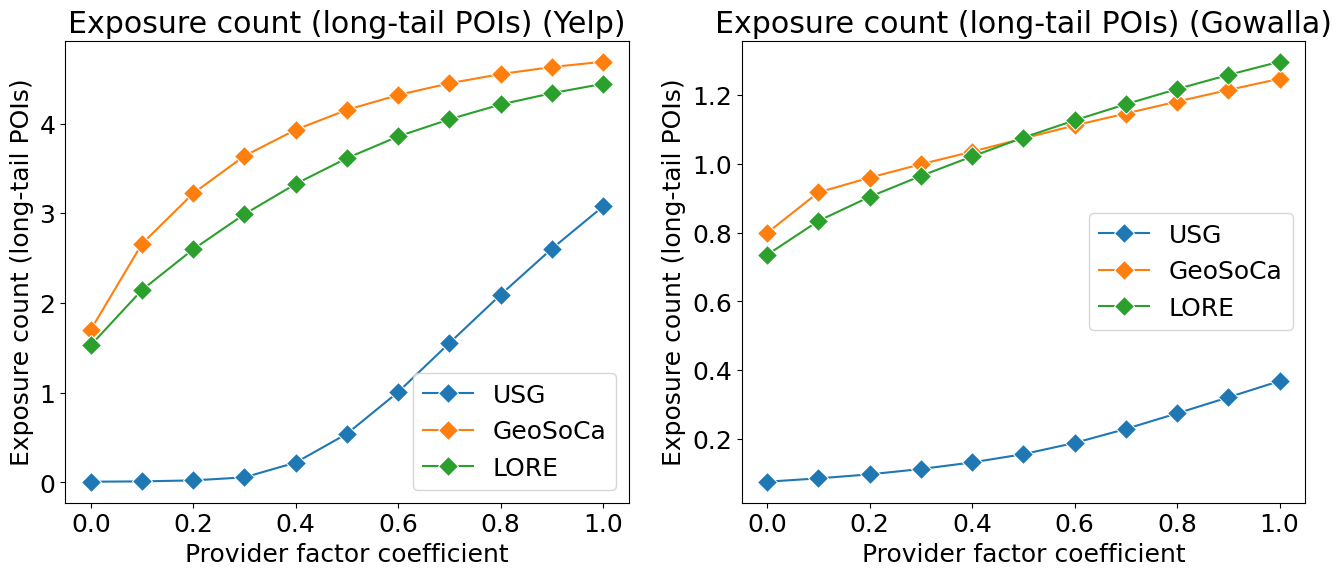}
            \caption{Long-tail Exposure v.s. Provider fairness factor $\alpha$}
            \label{fig:exp-longtail-provfactor}
            \vspace{-5px}
\end{figure}

\begin{figure}[ht!]
 \centering
            \includegraphics[width=0.90\linewidth]{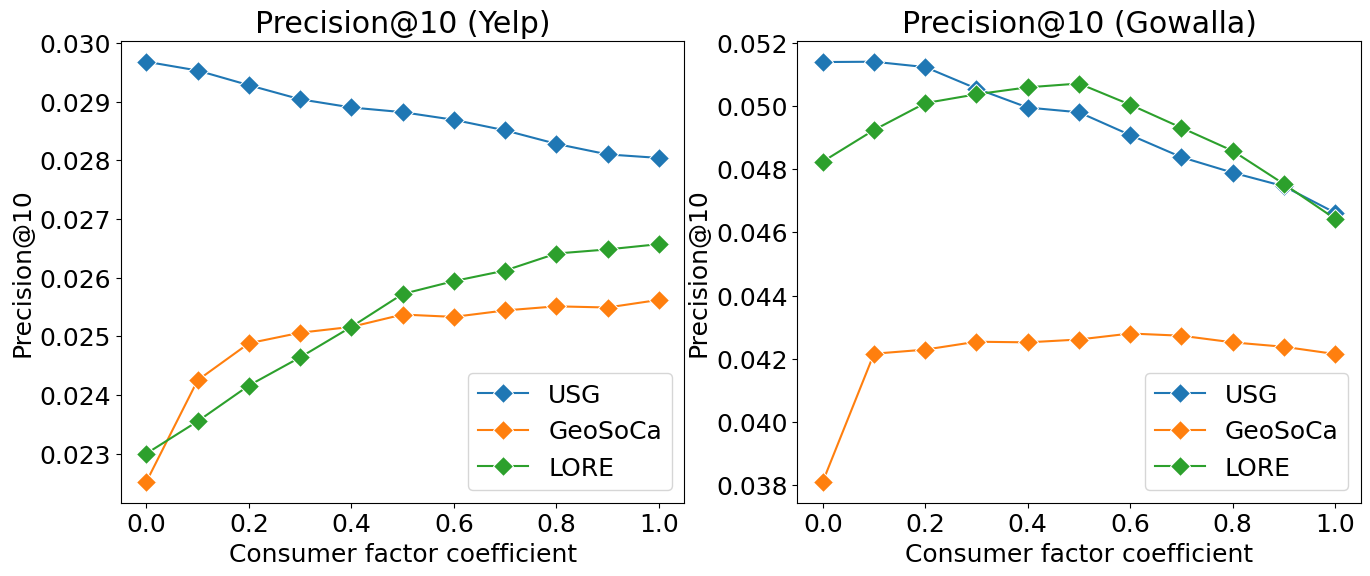}
            \caption{Precision@10 v.s. Consumer fairness factor $\beta$}
            \label{fig:prec-consfactor}
            \vspace{-5px}
\end{figure}

\begin{figure}[ht!]
\hspace*{\fill}%
     \centering
     \begin{subfigure}[T]{0.495\linewidth}
         \centering
            \includegraphics[width=\textwidth]{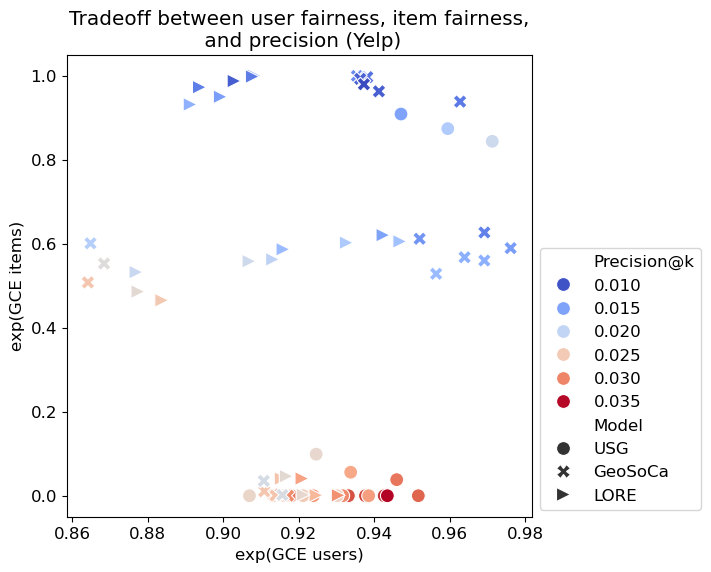}
     \end{subfigure}
     \hfill 
     \begin{subfigure}[T]{0.495\linewidth}
         \centering
            \includegraphics[width=\textwidth]{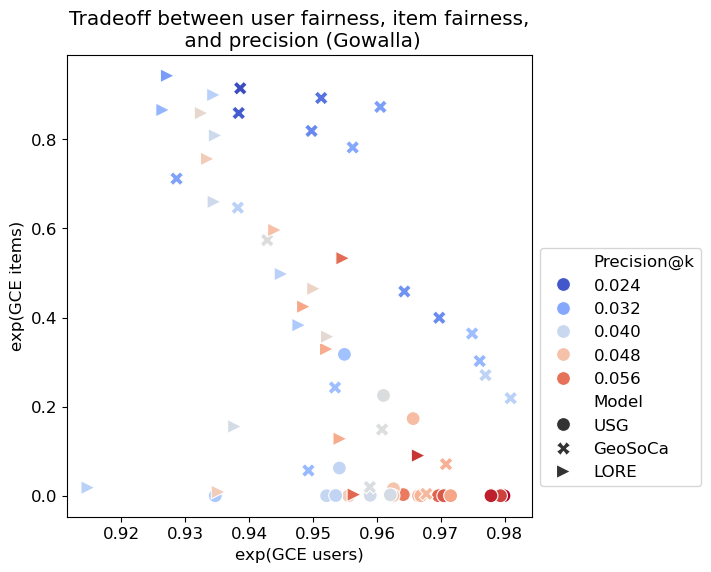}
        \end{subfigure}
        \hspace*{\fill}%
    \vspace{-5px}
    \caption{Scatterplot showing the tradeoff between GCE for users, GCE for items, and precision}
    \label{fig:tradeoff-scatter}
\end{figure}

For precision, the key performance metric, the linear exposure model consistently outperforms others, as indicated in bold in the results. This model's relative leniency on moderately popular POIs is illustrated by the curves in Figure \ref{fig:different-models} in the appendix. Conversely, the logistic model boosts exposure for the least popular POIs. The power law model, however, underperforms for the GeoSoCa and LORE models, particularly at higher $\alpha$ weights. Differences between the exposure models validate \textbf{H1} that the choice of exposure model is crucial. Specifically, the precision of the linear model is statistically significantly greater than that of the power law model. Pairwise comparisons between the linear and logistic models and logistic and power law models show no significant differences. Therefore, the linear model is chosen for the provider fairness factor due to its superior performance. We acknowledge this choice’s limitations and will include a comprehensive table of all pairwise results across models in Appendix \ref{sec:moremodeltests} for detailed comparison.
\vspace{1.5mm}\\
\textbf{Impact of the provider fairness factor}\\
After adopting the linear model as the provider fairness factor, we observed its effects on performance metrics related to \textbf{H2}. As shown in Table \ref{table:prec-ltexp-exposuremodel-yelp}, increasing $\alpha$ reduces precision but significantly enhances long-tail exposure. For instance, in the USG model, where baseline exposure is nearly zero, the increase in $\alpha$ substantially boosts exposure, as detailed in Figure \ref{fig:exp-longtail-provfactor}. This supports \textbf{H2}, demonstrating that provider fairness can enhance visibility for underrepresented POIs without overly compromising precision.
\\
\textbf{Impact of the consumer fairness factor}\\
Increasing the weight of the consumer fairness factor, denoted as $\beta$, aims to enhance precision for inactive users, aligning with \textbf{H3}. As depicted in Figure \ref{fig:prec-consfactor}, adjusting $\beta$ in the Yelp dataset initially improves precision for the GeoSoCa and LORE models from 2.3\% to 2.6\%. However, further increases cause precision to decline, notably in the USG model, where it drops from 3.0\% to 2.8\%. A Wilcoxon signed rank test confirms significant changes in precision across models with minor adjustments in $\beta$, underscoring the impact of fairness modifications on recommender systems.
\\
\textbf{Tradeoff between provider and consumer fairness}\\
Exploring the impact of combining provider and consumer fairness factors, as outlined in \textbf{H3}, can be seen through GCE metrics for different tradeoffs, detailed in Figure \ref{fig:tradeoff-scatter} and Table \ref{table:metrics-with-tradeoffs-yelp}. This figure plots the tradeoff between user precision GCE and item exposure GCE metrics, using color to denote precision levels and shapes to differentiate between models and $k$ values. Across the Yelp and Gowalla datasets, the Pareto front typically occupies the upper right region of each plot, highlighting a significant tradeoff where high user GCE often comes at the expense of item GCE, especially evident in the USG model which shows optimal precision. Table \ref{table:metrics-with-tradeoffs-yelp} further reveals that while GeoSoCa and LORE achieve the best precision with full weight on consumer fairness, this approach diminishes provider fairness. Conversely, the USG model demonstrates better performance with balanced fairness weights. Such adjustments to fairness weights reveal nuanced impacts on mean median distances, generally enhancing provider fairness across all models. This analysis underscores how specific properties of each model and dataset characteristics influence the effectiveness of combined fairness strategies, thus affecting overall outcomes in POI recommendation systems. For more details, see Appendix \ref{sec:moretrade}.

%% file: sections/conclusion.tex
To integrate consumer and provider fairness into pre-existing POI recommendation models, a post-filter approach was employed using 2 fairness factors added into the pre-filter score with varying weights. The provider factor gives higher scores for less popular POIs, using a linear model fit to the popularity distribution, so moderately popular recommendations are not severely penalized. The consumer factor attempted to increase recommendation quality for inactive users by recommending more popular POIs near previously visited ones. The results show that increasing the provider factor's weight resulted in improved exposure to the long-tail, up to 3-fold in the Yelp dataset, especially for USG where exposure was almost none. This came at the cost of lower precision, varying between models.
In some cases, the consumer factor improved precision but failed to enhance performance for inactive users. Nevertheless, combining both factors can yield reasonable results. Overall, there is a tradeoff between these factors, with the provider fairness factor achieving better long-tail exposure with minimal precision decreases for models like USG. Unfortunately, when comparing GCE metrics across models, a distinct correlation exists between low item exposure GCE values and higher precision, although this depends on the dataset, as the LORE model reached a middle ground between the metrics in the Gowalla dataset.

In the future, we plan to expand our research to include various models and strategies for consumer fairness and other objectives. We will explore different splitting methods and refine our accuracy assessments, aiming to comprehensively address complex fairness issues and enhance POI recommender systems' effectiveness.

%% file: sections/aux.tex
\subsection{Evaluation Metrics}
For evaluation, we will primarily use Precision@k as our accuracy metric, as this is the most commonly used one in the field. Precision@k simply checks if the target POI to be predicted lies in the first $k$ entries in the recommendation list, which would suffice for a use case where hits (i.e. visits to a POI) are sparse.

We can measure precision and exposure across the user and item groups to evaluate user and item fairness. For a single measurable metric of fairness, the Generalized Cross-Entropy (GCE) of the metric distribution can be used for both cases \cite{p31-rahmani2022unfairness} \cite{p55-deldjoo2021flexible}. For categorical attributes $a \in A$ with respect to which fairness will be computed, the GCE for a given model $m$ is defined as such:

\begin{equation}
    GCE(m, a) = \frac{1}{\beta \cdot (1 - \beta)}\Bigg[\sum_{a_j} p^\beta_f(a_j) p^{(1-\beta)}_m(a_j) - 1\Bigg]
\end{equation}

Here, $p_m$ and $p_f$ represent the metric distributions of the model outputs and a theoretical fair model, respectively. The index $\beta$ represents the index of the unfairness measure being used, which includes the Hellinger distance ($\beta = \frac{1}{2}$), Pearson's $\chi^2$ ($\beta = 2$), Kullback-Leibler divergence ($\lim_{\beta \rightarrow 1}$), and a few others. As mentioned in \cite{p55-deldjoo2021flexible}, the Pearson's $\chi^2$ measure is more robust to outliers, so a value of $\beta = 2$ will be used.

To adapt this GCE formulation to consumer fairness, we define a recommendation gain for each user based on the lists recommended to the user. Each of the POIs has an associated gain to them. In the case of recommendation hits or relevance, we can add 1 for each POI in each recommendation result that corresponds to one visited by the user. We can define this similarly to the attention function used in calculating item exposure $a(p, P_R)$. Suppose we define $\phi(u, p)$ as an indicator function that equals either 0 or 1 depending on whether or not the user has visited the POI. In that case, we can define the recommendation gain, as well as the overall metric distribution as follows:

\begin{equation}
    p_m(a_i) = \frac{1}{Z} \sum_{u \in U_{a_i}} rg_u = \frac{1}{Z} \sum_{u \in U_{a_i}} \sum_{P_{Ru} \subset P} \sum_{p \in P_{Ru}} \phi(u, p) \cdot a(p, P_{Ru})
\end{equation}

Note that $Z$ is a simple normalization factor which ensures that the sum of the metric distribution values equals 1. We can define a similar metric for provider fairness. We tweak the indicator function to indicate if a POI is in a recommendation list, and restate it as $\delta(p, P_{Ru})$. The resulting recommendation gain for POIs is as follows:

\begin{equation}
    p_m(a_i) = \frac{1}{Z} \sum_{i \in I_{a_i}} rg_i = \frac{1}{Z} \sum_{i \in I_{a_i}} \sum_{P_{Ru} \subset P} \delta(p, P_{Ru})
\end{equation}

Because of the constraints of the POI recommendation task, metrics with respect to these constraints can also be defined. For geographical distance, a simple average distance of recommended POIs to users may be employed.

\subsection{Modelling the provider fairness factor using various functions}

\begin{figure}[hbt!]
    \centering
    \includegraphics[width=0.3\textwidth]{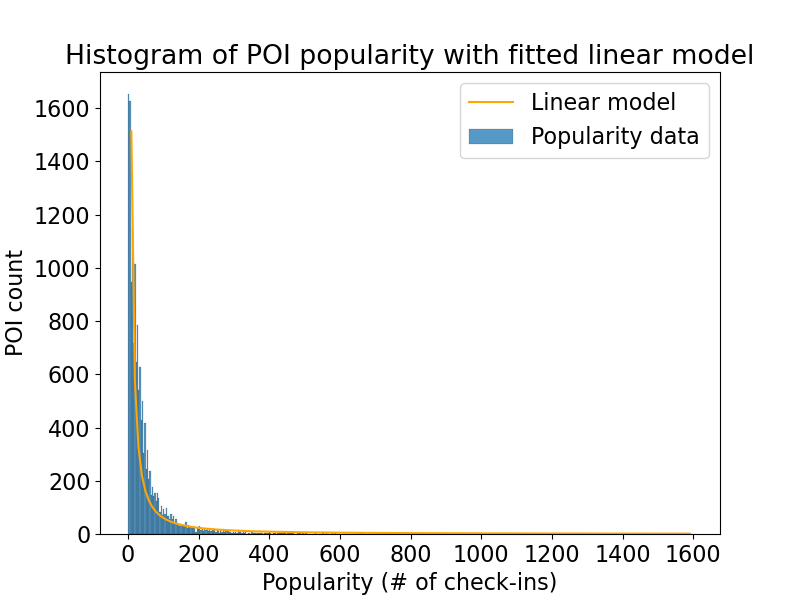}
    \caption{Histogram of popularity /check-in counts in the Yelp training dataset, with ridge regression linear model, $\alpha = 10.0$. \\}
    \label{fig:yelp-histmodel}
\end{figure}

\begin{figure}[hbt!]
    \centering
    \includegraphics[width=0.3\textwidth]{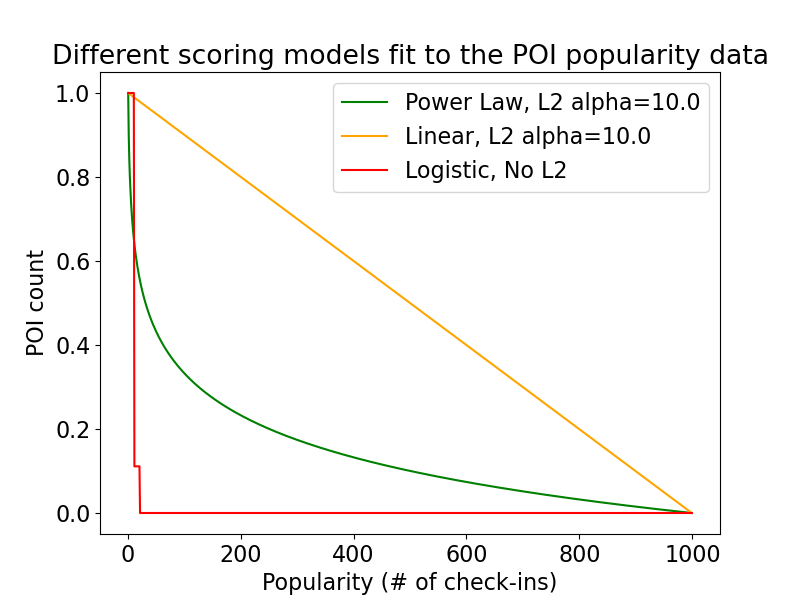}
    \caption{Different provider fairness scoring models.  \\}
    \label{fig:different-models}
\end{figure}

As mentioned above, the popularity of POIs follows a power law distribution, which can be modelled with a power law, where the score of a POI is inversely proportional to its check-in count, helping to elevate less popular POIs. The histogram of popularity values are shown in Figure  \ref{fig:yelp-histmodel}, which includes the power law model determined via a linear regression on the log-log transformed data. In addition to the power law, we also use linear and logistic models that represent different methods of scaling the fairness factor to popularity. These are also plotted in Figure \ref{fig:different-models}.



\subsection{Mann-Whitney U tests for the provider fairness scoring models}\label{sec:moremodeltests}

Aside from the Kruskal-Wallis tests ran across the 3 scoring models, we also ran pairwise Mann-Whitney U tests. The significance results are shown below in Figure \ref{fig:mannwhitney}, referencing the values in Table \ref{table:prec-ltexp-exposuremodel-yelp}. Note that the differences in long-term exposure are immediately significant even for low factor weights. The significance in differences in precision come a little bit later with higher factor weights, specifically in the Yelp dataset.

\begin{figure}[ht]
\centering
\addtolength{\tabcolsep}{-0.2em}
\resizebox{\linewidth}{!}{\begin{tabular}{|c c|c c c c |c c c c |}
    \hline
    \multicolumn{2}{|c|}{\textbf{Yelp Dataset}}&\multicolumn{4}{c|}{\textbf{Precision}}&\multicolumn{4}{c|}{\textbf{L-T Exposure}} \\
    Model & Comparison & 0.25 & 0.50 & 0.75 & 1  & 0.25 & 0.50 & 0.75 & 1 \\
    \hline\hline
     GeoSoCa & Lin. v.s. Power. &  &  & \checkmark & \checkmark & \checkmark & \checkmark & \checkmark & \checkmark \\
      & Power. v.s. Log. & \checkmark & \checkmark & \checkmark &  & \checkmark & \checkmark & \checkmark &  \\
      & Lin. v.s. Log. & \checkmark & \checkmark & \checkmark & \checkmark & \checkmark & \checkmark & \checkmark & \checkmark \\
    \hline\hline
     LORE & Lin. v.s. Power. & \checkmark & \checkmark & \checkmark & \checkmark & \checkmark & \checkmark & \checkmark & \checkmark \\
      & Power. v.s. Log. &  & \checkmark & \checkmark &  & \checkmark & \checkmark &  & \checkmark \\
      & Lin. v.s. Log. &  & \checkmark & \checkmark & \checkmark & \checkmark & \checkmark & \checkmark & \checkmark \\
    \hline\hline
     USG & Lin. v.s. Power. &  & \checkmark & \checkmark & \checkmark & \checkmark & \checkmark & \checkmark & \checkmark \\
      & Power. v.s. Log. &  & \checkmark & \checkmark &  & \checkmark & \checkmark & \checkmark & \checkmark \\
      & Lin. v.s. Log. &  &  & \checkmark & \checkmark & \checkmark & \checkmark & \checkmark & \checkmark \\
    \hline
\end{tabular}}
\resizebox{\linewidth}{!}{\begin{tabular}{|c c|c c c c |c c c c |}
    \hline
    \multicolumn{2}{|c|}{\textbf{Gowalla Dataset}}&\multicolumn{4}{c|}{\textbf{Precision}}&\multicolumn{4}{c|}{\textbf{L-T Exposure}} \\
    Model & Comparison & 0.25 & 0.50 & 0.75 & 1  & 0.25 & 0.50 & 0.75 & 1 \\
    \hline\hline
     GeoSoCa & Lin. v.s. Power. & \checkmark & \checkmark & \checkmark & \checkmark &  & \checkmark & \checkmark & \checkmark \\
      & Power. v.s. Log. & \checkmark & \checkmark & \checkmark & \checkmark & \checkmark & \checkmark & \checkmark & \checkmark \\
      & Lin. v.s. Log. & \checkmark & \checkmark & \checkmark & \checkmark & \checkmark & \checkmark & \checkmark & \checkmark \\
    \hline\hline
     LORE & Lin. v.s. Power. & \checkmark & \checkmark & \checkmark & \checkmark & \checkmark & \checkmark & \checkmark & \checkmark \\
      & Power. v.s. Log. & \checkmark & \checkmark & \checkmark & \checkmark & \checkmark & \checkmark & \checkmark & \checkmark \\
      & Lin. v.s. Log. & \checkmark & \checkmark & \checkmark & \checkmark & \checkmark & \checkmark & \checkmark & \checkmark \\
    \hline\hline
     USG & Lin. v.s. Power. & \checkmark &  & \checkmark & \checkmark & \checkmark &  & \checkmark & \checkmark \\
      & Power. v.s. Log. &  & \checkmark & \checkmark & \checkmark & \checkmark & \checkmark & \checkmark & \checkmark \\
      & Lin. v.s. Log. &  & \checkmark & \checkmark & \checkmark & \checkmark & \checkmark & \checkmark & \checkmark \\
    \hline
\end{tabular}}
\caption{Significance results of pairwise Mann-Whitney U tests for different provider fairness scoring models.}
\label{fig:mannwhitney}
\end{figure}

\subsection{More results on the tradeoff between provider and consumer fairness}\label{sec:moretrade}

Now that we have the 2 fairness factors to consider, the next question is to see how a combination of the 2 will affect the same metrics. For the graphs, a weight of 1.0 is shared between the consumer and provider factors. Although we see that even a high consumer fairness factor will decrease accuracy metrics for the USG model, it still affects it less than a high provider fairness factor. For GeoSoCa and LORE in particular, the performance increases as more weight is given to consumer fairness, and away from producer fairness. USG shows a peak midway, at a weight distribution of $\alpha = 0.3, \beta = 0.7$ for the Yelp dataset, and $\alpha = 0.2, \beta = 0.8$ for the Gowalla dataset.

\begin{figure}[!htb]
    \centering
    \includegraphics[width=\linewidth]{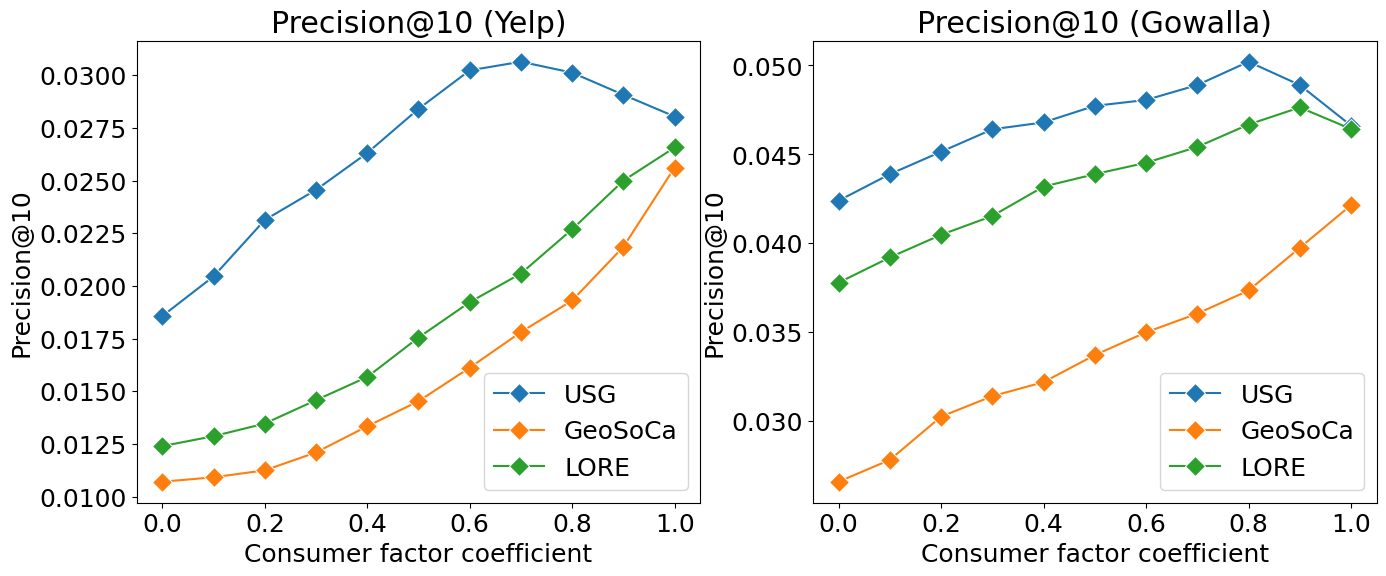}
    \caption{Precision@10 v.s. Tradeoff between consumer and provider fairness factors}
    \label{fig:prec-bothfactor}
\end{figure}

When looking at the fairness metrics for both long-tail items and inactive users, we see the expected changes, where an increase in the consumer fairness factor's weight corresponds to a decrease in long-tail exposure, and an increase in precision for inactive users. In particular, even a small addition of consumer fairness weight can cause a large decrease in exposure, especially for USG, where a weight of $\beta = 0.2$ can cause a 3-fold decrease in long-tail exposure.

\begin{figure}[!htb]
    \centering
            \includegraphics[width=\linewidth]{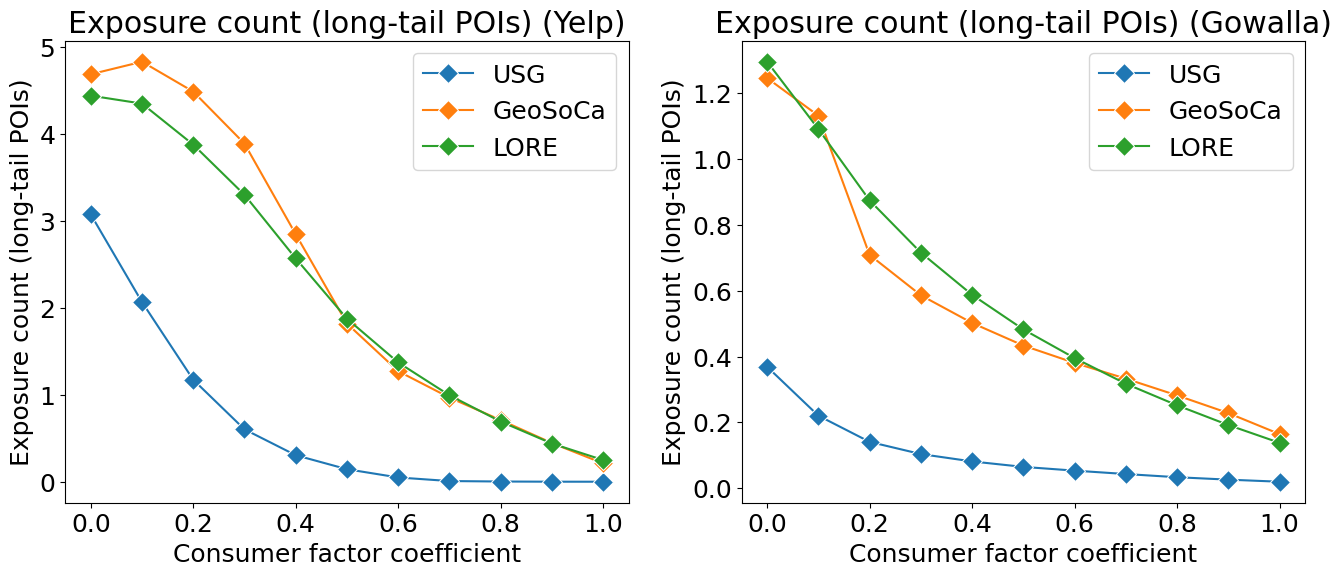}
            \caption{Exposure for Long tail items (bottom 80\%)}
            \label{fig:explt-bothfactor}
\end{figure}

\begin{figure}[!htb]
    \centering
            \includegraphics[width=\linewidth]{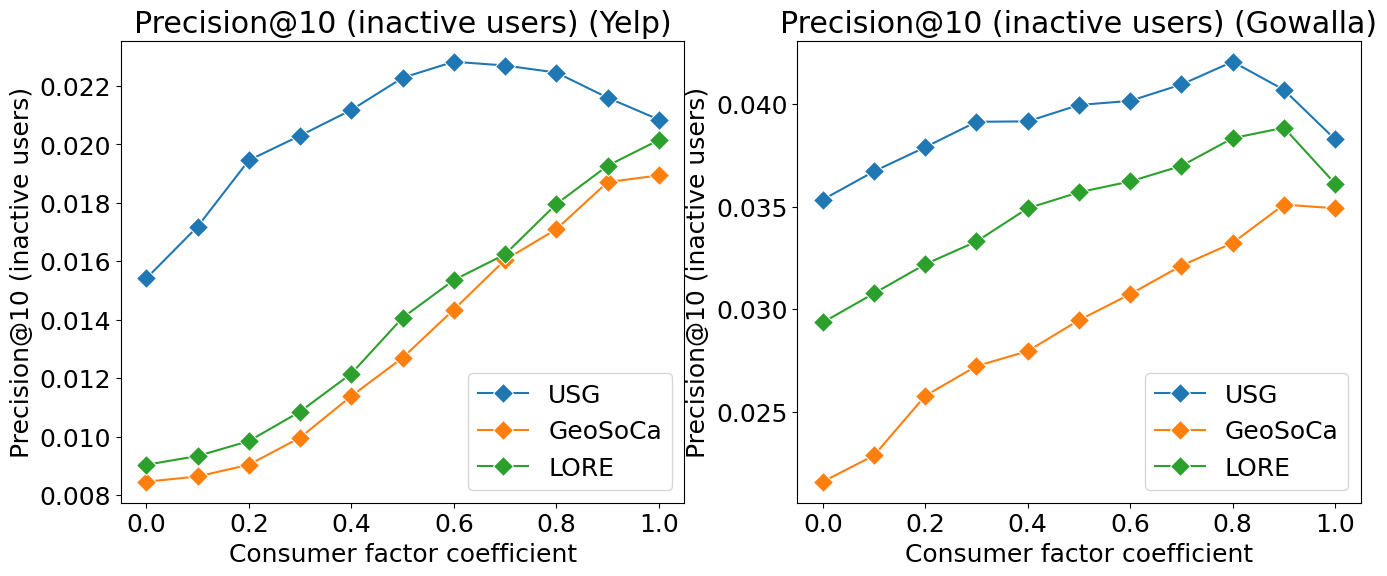}
            \caption{Precision@10 for inactive users (bottom 80\%)}
            \label{fig:precinactive-bothfactor}
\end{figure}

\subsection{Dataset statistics}

The table below displays the details about the 2 datasets used in the study, Yelp and Gowalla. Note that the sparsity is below even 1\%, highlighting the sparsity of this kind of recommendation task.

\begin{figure}[ht]
\centering
\begin{tabular}{|p{3cm}|c|c|}
    \hline
    \textbf{Dataset}& \textbf{Yelp} & \textbf{Gowalla} \\
    \hline
    Users & 7,135 & 5,628 \\
    \hline
    POIs & 16,621 & 31,803 \\
    \hline
    Checkins & 774,320 & 483,846 \\
    \hline
    Sparsity & 0.65\% & 0.27\% \\
    \hline
    Active/Inactive Users & 1,427 / 5,708 & 1,125 / 4,503 \\
    \hline
    Long-tail/Short-head POIs & 12,413 / 3,162 & 24,700 / 6,243 \\
    \hline
    Long-tail/Short-head Check-ins & 280,255 / 494,065 & 164,459 / 319387 \\
    \hline
\end{tabular}
\caption{Details about the datasets used.}
\label{table:dataset-stats}
\end{figure}



